# Isotropic Quantum Scattering and Unconventional Superconductivity

T. Park[1,2], V. A. Sidorov[1,3], F. Ronning[1], J.-X. Zhu[1], Y. Tokiwa[1], H. Lee[1], E. D. Bauer[1], R. Movshovich[1], J. L. Sarrao[1], and J. D. Thompson[1]

[1] *Los Alamos National Laboratory, Los Alamos, NM 87545, USA*

[2] *Department of Physics, Sungkyunkwan University, Suwon 440-746, South Korea*

[3] *Vereshchagin Institute of High Pressure Physics, RAS, 142190 Troitsk, Russia*

**Superconductivity without phonons has been proposed for strongly correlated electron materials that are tuned close to a zero-temperature magnetic instability of itinerant charge carriers[1]. Near this boundary, quantum fluctuations of magnetic degrees of freedom assume the role of phonons in conventional superconductors, creating an attractive interaction that "glues" electrons into superconducting pairs. Here we show that superconductivity can arise from a very different spectrum of fluctuations associated with a local or Kondo-breakdown quantum-critical point[2-5] that is revealed in isotropic scattering of charge carriers and a sub-linear temperature-dependent electrical resistivity. At this critical point, accessed by applying pressure to the strongly correlated, local-moment antiferromagnet $CeRhIn_5$, magnetic and charge fluctuations coexist and produce electronic scattering that is maximal at the optimal pressure for superconductivity. This previously unanticipated source of pairing glue[6] opens possibilities for understanding and discovering new unconventional forms of superconductivity.**

In conventional superconductors, excitations of a solid's crystal lattice provide an attractive interaction that binds itinerant electrons into pairs with zero net spin and momentum[7]. The resulting state of superconductivity is very sensitive to the presence of paramagnetic impurities, such as cerium: coupling of the paramagnetic spin *S* with the spin *s* of itinerant electrons, in the form of $J\mathbf{S} \cdot \mathbf{s}$, scatters electrons and breaks electron pairs[8]. Consequently, it has been a challenge to explain discoveries of unconventional superconductivity in strongly correlated electron materials with a dense array of paramagnetic ions, such as Ce-based heavy-fermion compounds[9]. This same form of spin-spin coupling is the origin of the Kondo effect which gives itinerant charge carriers their heavy mass in heavy-fermion systems and, in the process, creates a 'large' Fermi volume that consumes the paramagnetic electrons, e.g., the 4f electrons of Ce (ref. 10). The proximity of superconductivity in heavy fermions, such as the prototype $CeCu_2Si_2$, to a spin instability of their large Fermi surface suggests that magnetic fluctuations provide the attractive "glue" in the same way that lattice fluctuations do in conventional superconductors[1,9]. As a spin-density instability is tuned by a non-thermal parameter, such as pressure, toward a continuous $T$=0 K transition, magnetic excitations become quantum mechanically critical, leading to a divergent magnetic susceptibility at the ordering wave vector **Q**, a condition favourable for creating superconducting electron pairs with finite angular momentum[9]. Quantum fluctuations of the spin density also strongly scatter electrons on parts of the Fermi surface connected by the wave vector **Q**, giving an electrical resistivity that increases from $T$=0 K with a power-law dependence $\rho \sim T^\varepsilon$, where $1 \leq \varepsilon < 3/2$, which is distinct from the characteristic $T^2$ dependence of a Fermi liquid (FL) that emerges as the strongly correlated electron system is tuned away from its quantum-critical point[11].

Quantum criticality of a very different kind has been proposed for heavy-fermion materials whose magnetism derives from the indirect interaction among localized magnetic moments[2-5]. Deep in the magnetically ordered phase, the Kondo effect is absent and the Fermi volume is small, i.e., does not include the localized electrons, but, as the $T$=0 paramagnetic/localized magnetic boundary is approached, the Fermi volume jumps in size. A consequence of such a quantum-phase transition is that the Fermi surface becomes critical at the $T$=0 magnetic/paramagnetic boundary, consistent with a spatially local character of the criticality and creating a spectrum of fluctuations that dominate physical properties, such as resistivity, even at temperatures well above $T$=0 K (ref. 12). Compared to a spin-density quantum-phase transition that affects only a portion of the Fermi surface and involves only magnetic degrees of freedom, a local-type of quantum-phase transition is more 'violent', producing fluctuations in spin and charge channels that separately or collectively could promote an attractive interaction between electron pairs. However, the lack of a theoretical basis for superconductivity and the absence of superconductivity in the prime candidates for local criticality, $YbRh_2Si_2$ and $CeCu_{6-x}Au_x$, question the viability of superconductivity mediated by locally critical fluctuations[6]. Using the heavy-fermion antiferromagnet $CeRhIn_5$ as an example, we report the first evidence that fluctuations from a local form of quantum criticality can provide a new route to unconventional superconductivity.

Figure 1 displays a temperature-pressure map of the local exponent $\varepsilon$ ($\equiv \partial \ln(\rho(T)-\rho(T=0)) / \partial \ln T$) of the electrical resistivity of $CeRhIn_5$. Funnel-shaped yellow boundaries in Fig. 1 that emerge near the maximum pressure-induced superconducting transition temperature $T_c$ demarcate a lower crossover temperature from an unusual state with a sub-T-linear dependent resistivity, $\rho \sim T^{0.85}$. (Representative pressure- and temperature-dependent resistivity curves and their analysis, shown in Supplementary Figs. S1 and S2, document the origin of the colour map and phase boundaries.) In the low pressure limit, the yellow boundary reflects the onset of short-ranged antiferromagnetic spin correlations over a small interval above the Néel temperature $T_N$ (ref. 13), but signals a change in resistivity to a $T^{3/2}$ dependence before settling into a heavy FL state at $T < T_{FL}$ in the high-pressure limit. The upper boundary of this unusual state is determined by $T^{max}/2$, where $T^{max}$ is the point where the resistivity is a maximum. At 2.35 GPa, where there is no magnetic order, this phase with sublinear resistivity extends to progressively lower temperatures as superconductivity is suppressed by an applied field. At 10 T, which is slightly higher than $H_{c2}(0)$, the field necessary to suppress superconductivity completely, the weakly field-dependent sub-T-linear resistivity holds from ~10 K down to 250 mK, and crosses over to a $T^2$ FL dependence below 150 mK (see Supplementary figure S3). We emphasize that this sub-$T$-linear behaviour with a single temperature exponent over a broad temperature range is distinct from crossover behaviour that is often observed in heavy fermion compounds as they are warmed from a low temperature Fermi-liquid $T^2$ behaviour to a resistivity maximum near 100 K: the temperature exponent in the crossover regime varies with temperature, i.e., there is no unique power-law behaviour.

Specific heat studies as a function of pressure have revealed that the Néel transition extends continuously to $T$=0 at 2.35 GPa as a magnetic field suppresses superconductivity[14]. This magnetic quantum-critical point at 2.35 GPa (P2) offers a possible explanation for the strange metallic behaviour that emanates from the point of maximum $T_c$; however, the sub-$T$-linear resistivity is not anticipated by any theory of spin-density criticality[15]. The nonconformity of CeRhIn$_5$ to the spin-density scenario indicates that the nature of the quantum-critical point in CeRhIn$_5$ may be different from a magnetic instability of the Fermi surface. When coupled with quantum-oscillation measurements at 2.35 GPa (ref. 16), which show an abrupt change in the Fermi-surface, the convergence of multiple energy scales represented by the crossover and phase-transition boundaries in Fig. 1 is consistent with an interpretation of a local form of criticality that involves magnetic as well as fermionic degrees of freedom[2-6], where, now, the yellow boundary at high pressures can be understood as a crossover temperature, below which incoherent scattering due to quantum fluctuations starts to yield to coherent scattering among the lattice of Kondo sites.

Strong support for the spatially local nature of this criticality comes from electrical transport measured perpendicular ($\rho_{ab}$) and parallel to the c-axis ($\rho_c$) (see Fig 2). At temperatures higher than ~100 K, where charge scattering comes predominantly from randomly oriented Ce 4f moments, the ratio $\rho_{ab}/\rho_c \approx 0.5$ is essentially independent of pressure and set primarily by intrinsic crystalline anisotropy. On the other hand, below $T_{FL}$ at high-pressures, transport anisotropy increases by a factor of 2.5 to give $\rho_{ab}/\rho_c \approx 0.2$. We ascribe this increase in anisotropy to characteristics of the large Fermi volume that now includes the 'Kondo-ized' 4f-electrons of Ce. In the large Fermi-volume, sister superconductor CeCoIn$_5$, the effective mass anisotropy $m^*_a / m^*_c \approx 0.18$ (ref. 17), where $m^*$ reflects many-body correlation effects that also are present in CeRhIn$_5$ below $T_{FL}$. The pronounced anisotropy observed in the FL state becomes progressively less obvious with decreasing pressure. In the quantum-critical regime, where the resistivity is sub-linear in temperature, the anisotropy ratio is almost temperature independent and is similar to that at room temperature. With the assumption of independent scattering sources, the total scattering rate is $1/\tau = 1/\tau_i + 1/\tau_q$, where $\tau_i$ and $\tau_q$ are the collision times for scattering by impurities and by excitations, respectively. Potential scattering from impurities is isotropic, leaving crystal anisotropy, which changes very weakly with pressure[18], and scattering from fluctuations associated with the quantum phase transition as sources of resistive anisotropy. The absence of new anisotropy in the quantum-critical regime argues against a spin-density quantum-critical interpretation because, in this case, scattering from hot spots spanned by **Q** at the Fermi surface is expected to be highly anisotropic[19], reflecting the Fermi surface topology of the large-volume paramagnetic state. Instead, isotropic scattering induced over the entire Fermi surface at a local or Kondo-breakdown quantum-critical point is consistent with our observations. Model calculations, presented in the Supplementary figure S5, show that the sub-linear temperature dependence of electrical resistivity can arise from isotropic scattering from fluctuations associated with a local quantum-critical point. We note that a similar sub-T-linear resistivity has been shown to be present in a Kondo-breakdown model[5].

Could fluctuations from a local form of quantum criticality be responsible for the unconventional superconductivity of $CeRhIn_5$? Figure 3 displays a map of the isothermal resistivity, normalized by the resistivity in the normal metallic phase at 5.2 GPa, as a function of pressure. As seen in this figure and in Supplementary Fig. S6, the highest scattering rate and the highest $T_c$ occur simultaneously, which appears to contradict the conventional view that scattering is harmful to unconventional superconductivity[20, 21]: superconductivity disappears when the resistivity reaches a comparable value of 20 $\mu\Omega$ cm in disordered $CeCoIn_5$. Unlike chemical substitution, pressure does not induce extra disorder. Instead, the maximum resistivity in the vicinity of the quantum-critical point is due to the build up of quantum fluctuations that amplifies scattering by the small number of impurities in this very pure single crystal of $CeRhIn_5$. The high spectral density of quantum fluctuations, reflected in the maximum of $\rho_c(2.3K)$, then provides the glue for optimal electron pairing.

In $CeCu_2Si_2$, where two domes of superconductivity emerge with applied pressure, spin-density and valence fluctuations are proposed to mediate electron pairing separately for each dome, with a factor of four higher $T_c$ in the dome associated with valence fluctuations[22]. In $CeRhIn_5$, the local nature of quantum criticality implies simultaneous fluctuations in spin and charge channels. Which channels or channel dominate(s) the pairing interaction in $CeRhIn_5$ cannot be resolved by our experiments, but the multi-criticality of $CeRhIn_5$ appears to be a key to understanding the sub-$T$-linear resistivity and the robustness of the superconductivity over a wide pressure range. This work raises fundamental questions that need to be addressed: i) how do multiple fluctuating channels interplay with each other to produce superconductivity? ii) what is the appropriate description of fluctuations that arise from a local or Kondo-break down quantum-critical point? iii) why is there no superconductivity in comparably pure crystals of $YbRh_2Si_2$? and, iv) could a similar analysis be applicable to other heavy-fermion superconductors such as $PuCoGa_5$, where composite pairs of local moments and electrons are proposed[23] to condense to form superconductivity? Solution to these problems will guide the search for new examples of unconventional superconductivity and will be applicable broadly to other strongly correlated superconductors, such as the cuprate high-$T_c$ superconductors[24], where two domes of superconductivity or one extensive SC dome as a function of doping may naturally arise from multiple quantum-critical points.

**Acknowledgements** The authors thank Q. Si, C. D. Batista, A. V. Balatsky, C. Varma, Z. Nussivnov, D. Pines and N. J. Curro for helpful discussions. Work at Los Alamos National Laboratory was performed under the auspices of the US Department of Energy, Office of Science, with support from the Los Alamos Directed Research and Developmental program. VAS appreciates the support of RFBR (Grant 06-02-16590) and Program of the Presidium of RAS on Physics of Strongly Compressed Matter..

**Figure Legends**

Figure 1. Temperature versus pressure phase diagram of CeRhIn$_5$. Colours (right-hand bar) represent the local exponent $\varepsilon$ at zero magnetic field, defined as $\varepsilon = d\ln\Delta\rho / d\ln T$, where $\Delta\rho = \rho - \rho(T=0K) = AT^{\varepsilon}$. The resistivity $\rho$ was measured along the c-axis of this tetragonal crystal. The residual resistivity $\rho(T=0K)$ and temperature coefficient A were obtained with least-squares fitting to the power-law form at low temperatures and are shown in the Supplementary figure S1. Also shown are phase boundaries of local-moment antiferromagnetic (AF) order, superconductivity (SC), and the temperature below which the resistivity follows a $T^2$ temperature dependence characteristic of a Landau Fermi Liquid (FL). The cone-shaped region of green denotes a state of sub-$T$-linear resistivity, labelled NFL and discussed in the text, that appears to emanate from the dome of superconductivity (SC) where $T_c$ =2.3 K is a maximum. For comparison, the colours of the SC and AF phases are set to white and black, respectively.

Figure 2. Temperature-pressure variation of resistivity anisotropy. Colours (right-hand bar) describe $\rho_{ab}/\rho_c$, where $\rho_{ab}$ ($\rho_c$) is the resistivity measured perpendicular (parallel) to the tetragonal c-axis of CeRhIn$_5$. Apparent structure

in the colour map is due to an interpolation of data obtained at slightly different pressures used to determine $\rho_{ab}$ and $\rho_c$. Significantly, the resistivity anisotropy found at high temperatures persists to the lowest temperature in the same NFL pressure range where $\rho_c$ (Fig. 1) and $\rho_{ab}$ exhibit a sub-$T$-linear variation. AF, SC and FL denote antiferromagnetic, superconducting and Landau Fermi Liquid states, respectively. Representative data from which this map was constructed are shown in Supplementary figure S4. White colour is used to represent the SC phase.

Figure 3. Pressure-dependent c-axis resistivity. A colour contour map of the c-axis resistivity normalized by its value in the normal metallic state at 5.2 GPa, $\rho(P) / \rho(5.2\ GPa)$, is plotted in temperature-pressure plane. The colour key, right-hand bar, indicates this resistivity ratio is highest at temperatures and pressures where the resistivity is sub-$T$-linear (Fig. 1) and resistivity anisotropy (Fig. 2) is the same as at high temperatures. Maximum scattering appears above the maximum $T_c$ of the superconducting dome (SC). Representative isothermal cuts of the resistivity ratio, used to construct the colour map, are shown in Supplementary figure S6. White colour is used to represent the SC phase.

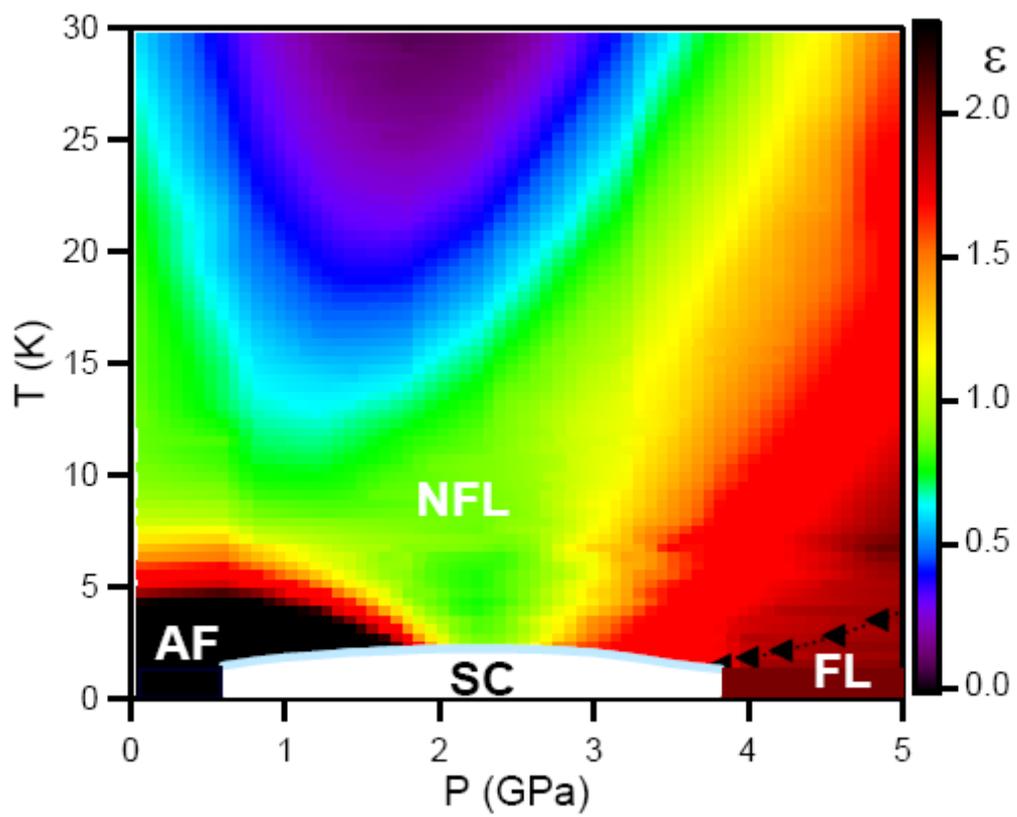

Figure 1

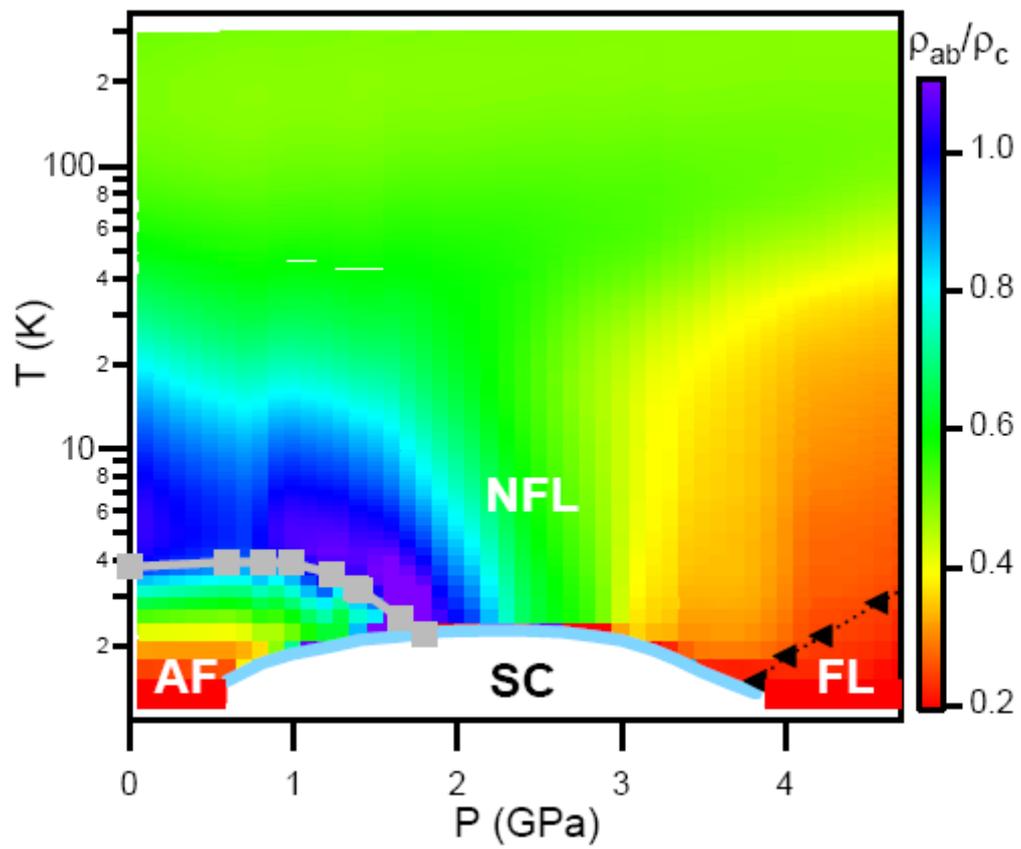

Figure2

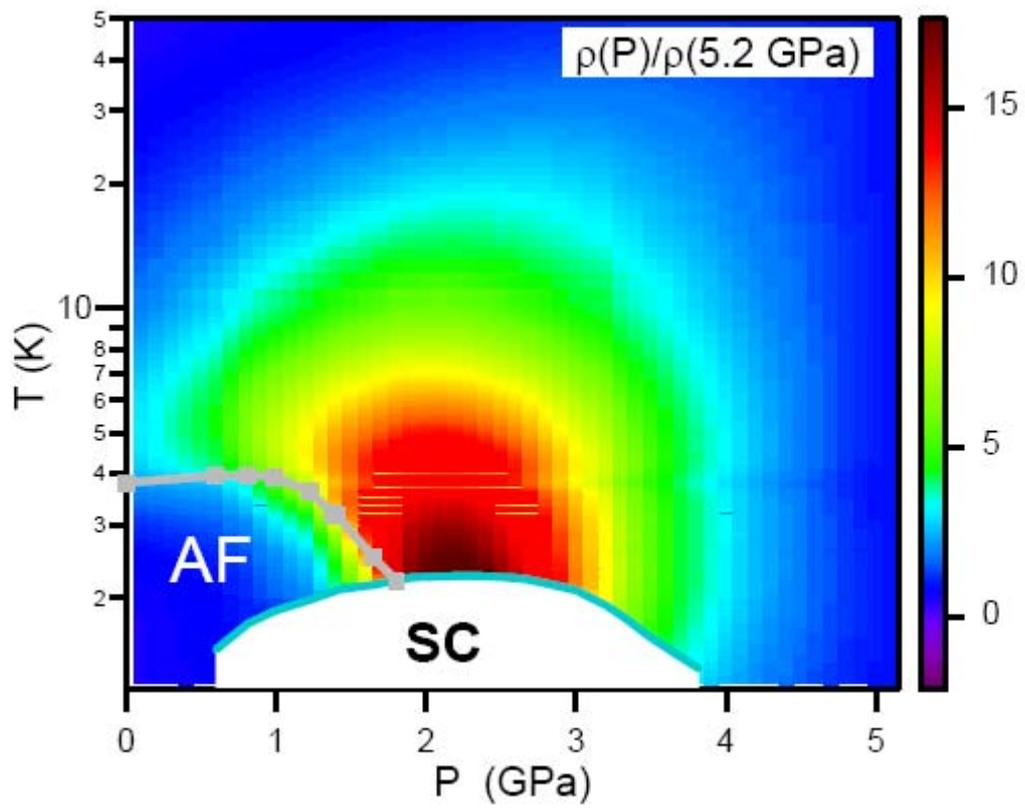

Figure 3

**Supplement Methods**

Single crystals of $CeRhIn_5$ and $LaRhIn_5$ were grown using excess In flux and screened by low field susceptibility measurements to ensure that no In flux remained on the surface or trapped inside the crystals. $CeRhIn_5$ crystallizes in the tetragonal $HoCoGa_5$ structure and becomes antiferromagnetic below 3.8 K, where the ordered magnetic moment of 0.8 $\mu_B$ is slightly reduced by the Kondo effect from the expected crystal-field doublet value of 0.92 $\mu_B$, indicating that $CeRhIn_5$ is in the local-moment limit[S1]. Electrical resistivity measurements were performed in a standard four-probe configuration, using spot-welded contacts. A residual resistivity ratio $\rho(300K)/\rho(T=0)$ of $CeRhIn_5$ of over 500 confirms the exceptional sample quality of the single crystals used in this work. Two different types of cells were used to generate pressure: a hybrid Be-Cu/NiCrAl clamp-type cell with silicone fluid as the pressure medium for pressures up to 2.5 GPa and a toroidal cell with a glycerol-water mixture as pressure medium for pressures up to 5.23 GPa. Hydrostatic conditions were indicated by a narrow (typically less than 15 mK) superconducting transition of a Sn (or Pb) manometer up to the highest pressure. In contrast to previous results of Ref S2, where a non-Fermi liquid state was present even at high-pressures and low temperatures, we found a $T^2$ dependence of the resistivity in comparable pressure-temperature ranges, which stresses the importance of hydrostatic pressure environments.

**Supplement Discussion**

Figures S1 and S2 show typical c-axis resistivity data and their analysis. From independent specific heat, ac susceptibility and neutron diffraction measurements[S1, S3, S4], features in curves plotted in Fig. S1a for pressures less than 2.12 GPa are known to be associated with long range antiferromagnetic order near 3 K or superconductivity at a lower temperature. Above these orders, there is a pressure-dependent temperature range where the resistivity follows a power-law form $\rho = \rho_0 + AT^\varepsilon$. Figure S1b shows that the temperature exponent $\varepsilon$ =0.85 (±0.02) over progressively larger temperature intervals above $T_c$ as pressure increases. We rule out possible superconducting fluctuations as a source of this sub-linear resistivity because the range over which it is found far exceeds any critical region associated with the mean-field-like transition at $T_c$ and because of its very weak field dependence. We also note that the resistivity at 2.12 GPa exhibits the sub-T-linear behaviour even without subtracting the small residual resistivity (see Fig. S1a) indicating that the power-law is not an artefact from subtracting an erroneous residual resistivity but is an intrinsic property. Values of the residual resistivity and temperature coefficient $A$ used in this panel are plotted in Fig. S2. At pressures greater than 3.7 GPa, the resistivity assumes a quadratic temperature dependence as shown in Fig. S1c. These representative data, taken from $\rho(T)$ measurements at over 20 roughly equally spaced pressures, were used to construct the colour map given in Fig. 1 of the text.

The effect of a magnetic field on the temperature exponent of resistivity is summarized in Fig. S3. When superconductivity is suppressed completely by a field of 10 T, the unusual exponent $\varepsilon$ decreases slightly from 0.83 (±0.02) to 0.71 (±0.02) and describes the temperature variation of resistivity from ~ 10 K to 250 mK at 2.35 GPa (Fig. S3b), before crossing over to a $T^2$ dependence at lower temperatures (Fig. S3c). Increasing the field to 14 T decreases the $T^2$ coefficient of resistivity and extends the temperature range over which $\rho \propto T^2$.

Representative resistivity data from which the colour map of anisotropy ratio was constructed (Fig. 2 in the text) are plotted in Fig. S4. In these data, the applied pressure for resistivity measurements parallel and perpendicular to the c-axis were nearly identical, ±0.05 GPa. In other cases, the difference in pressures was somewhat larger, up to ±0.2 GPa. To account for these pressure differences, data were interpolated to construct the ratio. Errors introduced by this interpolation are responsible for 'blockness' in the colour map, especially at lower temperatures where the resistivity ratio shows the strongest temperature dependence. For temperatures higher than 100 K, the resistivity of LaRhIn$_5$, the non-magnetic analog of CeRhIn$_5$, is less than 25% of that of CeRhIn$_5$, underscoring the dominance of Ce 4f spin exchange scattering over phononic contributions. In a relaxation-time approach, when the conduction electron velocity is assumed to be an averaged value over the Fermi surface, the transport anisotropy is essentially determined by $\dfrac{\rho_{ab}}{\rho_c} = \dfrac{\tau_c \int_{E_F} dS_c}{\tau_{ab} \int_{E_F} dS_{ab}}$, where $\tau_i$ and $dS_i$ are the relaxation time and projection of the Fermi surface element in the *i* direction, respectively.

The unusual sub-*T*-linear temperature dependence of the resistivity is not anticipated in any theory of spin-density quantum criticality, which gives at the weakest, a T-linear dependence[S5]. Assuming a *q* independent, isotropic scattering rate at a local quantum-critical point, the resistivity was calculated within the framework of the semiclassical Boltzmann equation [S5, S6]. In the local quantum-critical scenario, the momentum-dependent spin susceptibility for the *f*-electrons is modeled by $\chi(\mathbf{q};\omega) = 1/\left[I_q + \Lambda_0 (T/\Lambda)^\alpha \mathrm{M}(\omega/T)\right]$, where the spin exchange coupling $I_q = I_0 \left[\cos^2(q_x/2) + \cos^2(q_y/2)\right]$ and spin self-energy scaling function $\mathrm{M}(\omega/T) = (2\pi)^\alpha \exp\left[\alpha\psi(1/2 - i\omega/2\pi T)\right]$, with $\psi$ the digamma function. In the calculation, we chose realistic parameter values $\Lambda_0/2 = \Lambda = I_0 = 0.1$ and the scaling exponent $\alpha = 0.75$, found previously[S6] to be appropriate to local quantum criticality. As shown in Fig. S5, the calculated resistivity does assume a sub-linear temperature dependence as the conduction-band Fermi surface is tuned to be more isotropic. This tendency to a sub-T-linear dependence with increasing isotropy is consistent with studies of $CeRhIn_5$ that show isotropy within the tetragonal basal plane and at most a factor of 5 anisotropy between in-plane and out-of-plane susceptibility and resistivity. In a simplified hybridized band model, the transport properties come from the motion of conduction electrons and f-electrons. Our current calculation is focused on the critical region or very close to it, where the f electrons are localized and the contribution to conductivity is negligible. When f electrons become itinerant, the f component should not be neglected and our theoretical approach should be modified.

Examples of isothermal out-of-plane and in-plane normalized resistivity versus pressure curves are plotted in Fig. S6. For reference, the normalizing resistivity is taken to be that measured at the highest pressure of 5.2 GPa, which is the farthest removed from a phase boundary and the *T-P* region of unusual resistivity that arises from quantum-critical fluctuations. These data and those shown in the corresponding colour map (Fig. 3 of the text) emphasize strongly enhanced scattering in the pressure range where $T_c$ is a maximum.

**Supplementary Figures and Legends**

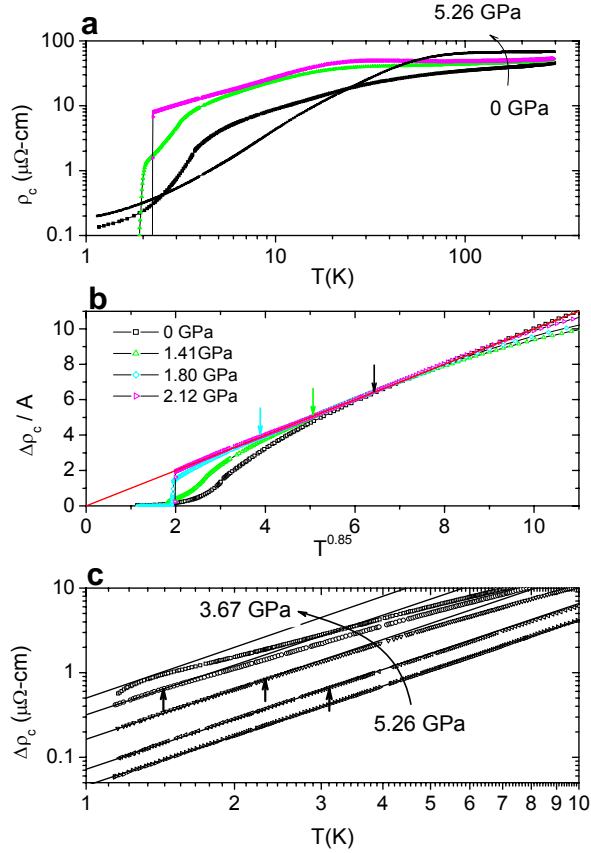

Figure S1. Representative c-axis resistivity data for CeRhIn$_5$ at selected pressures **a.** On a logarithmic scale, electrical resistivity of CeRhIn$_5$ for current along the c-axis. Each curve represents 0, 1.41, 2.12, and 5.26 GPa from the bottom to the top. **b.** Scaling behaviour of the c-axis resistivity. The normalized resistivity, $\Delta\rho / A$, is plotted against $T^{0.85}$, where $\Delta\rho = \rho - \rho_0 = AT^{0.85}$. The solid line is a reference for a $T^{0.85}$ dependence and arrows mark the temperature at which the resistivity deviates from the simple power-law form and constitutes the lower boundaries (yellow) of Fig. 1 in the main text. Fitting parameters obtained through the least-squares fitting scheme are displayed in the Supplementary figure S2. **c.** Temperature dependence of the c-axis resistivity at high pressures: 5.26, 4.84, 4.22, 3.82, 3.67 GPa from the bottom to the top. For pressures higher than 3.67 GPa, the electrical resistivity shows Landau-Fermi liquid behaviour, i.e., $\Delta\rho = AT^{1.96}$, where the temperature exponent is equal to 2 within experimental error. Solid lines are least squares fits to low-temperature data and arrows indicate the temperature where the resistivity deviates from the power-law form and constitutes $T_{FL}$ shown by solid triangles in Fig. 1 of the text.

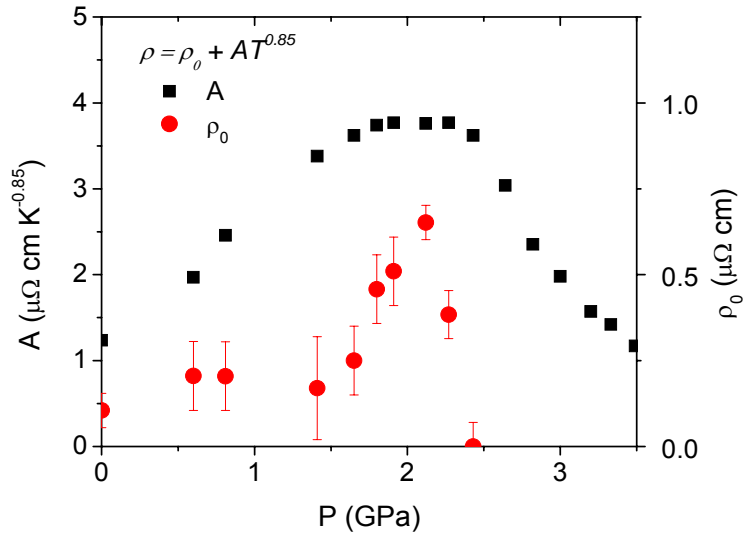

Figure S2. Pressure evolution of the resistivity fitting parameters. Fitting parameters A and $\rho_0$ are shown as a function of pressure, which were obtained from least-squares fits to the c-axis electrical resistivity of CeRhIn$_5$ in zero magnetic field, i.e., $\rho = \rho_0 + AT^{0.85}$. For pressures higher than 2.4 GPa, the small residual resistivity ($\rho_0$) could not be determined accurately because of the limited temperature range over which $\rho \propto T^{0.85}$. Both $\rho_0$ and $A$ are enhanced in the vicinity of the optimal pressure (2.1 GPa) where $T_c$ is the highest.

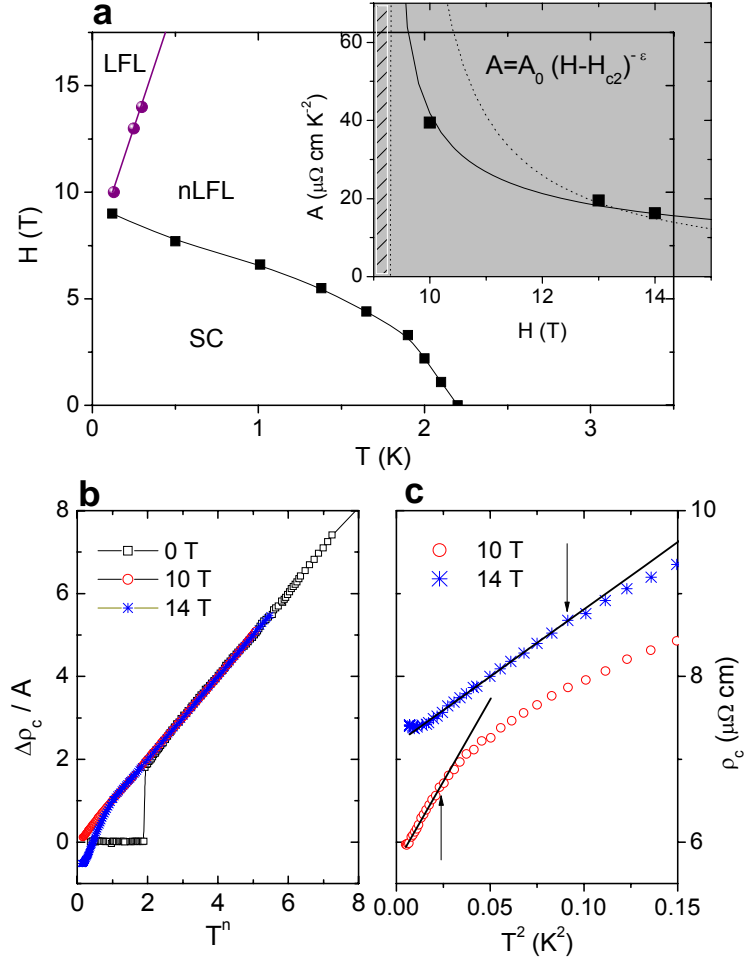

Figure S3. Summary of field effects on the resistivity of CeRhIn$_5$ at 2.35 GPa. **a**. Magnetic field versus temperature phase diagram of CeRhIn$_5$ at 2.35 GPa for field perpendicular to c-axis (**H** $\perp$ c). Inset: In the Fermi liquid regime, $\rho = \rho_0 + AT^2$, the $T^2$-resistivity coefficient $A$ is plotted against $H$. The solid (or dotted) line is for $n$ = 0.5 (or 1) and $H_{c2}$ is 9.3 T. **b** Scaling of the c-axis resistivity at 2.35 GPa at 0 (squares),10 (circles), and 14 T (crosses). Here $\varepsilon$ is 0.82, 0.71, and 0.86 for 0, 10, and 14 T, respectively, while the corresponding value for $\rho_0$ is 0, 5.2, and 9.4 $\mu\Omega\cdot$cm with $\Delta\rho = \rho - \rho_0 = AT^\varepsilon$. **c** The c-axis resistivity is plotted against $T^2$ for 10 (circles) and 14 T (squares). Solid lines are least squares fit of $\Delta\rho = AT^2$ and arrows mark the Fermi liquid temperature $T_{FL}$.

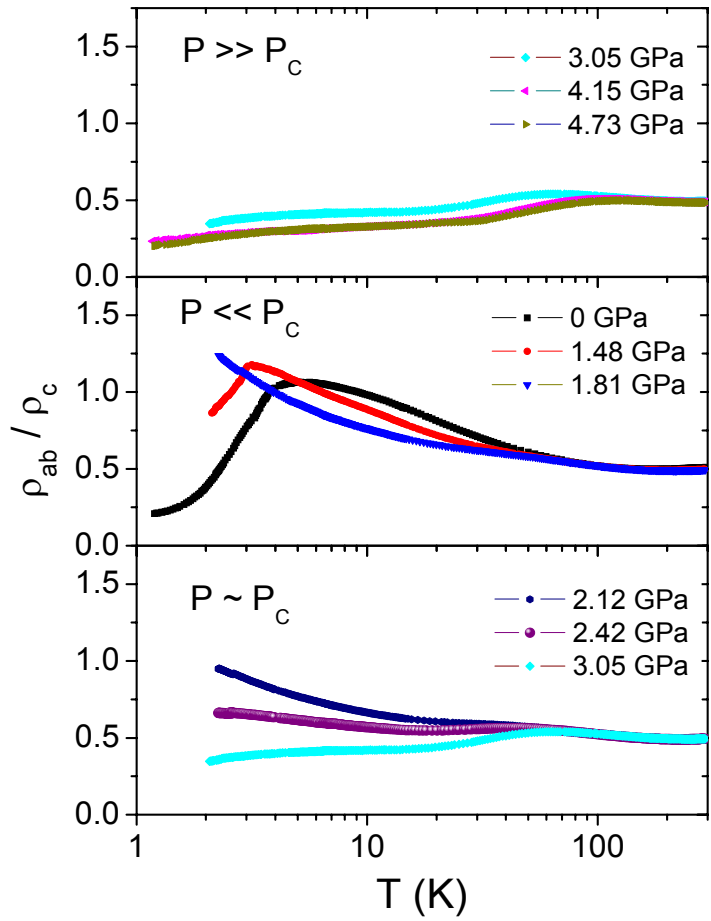

Figure S4. Isobaric cuts of the anisotropy in resistivity in CeRhIn$_5$. At a pressure of 2.42 GPa, the ratio is almost independent of temperature (bottom panel), while the anisotropy either decreases or increases with temperature for the low- (middle panel) or high-pressure limits (top panel), respectively. The broad peaks at low temperatures and low pressures reflect the presence of an antiferromagnetic transition.

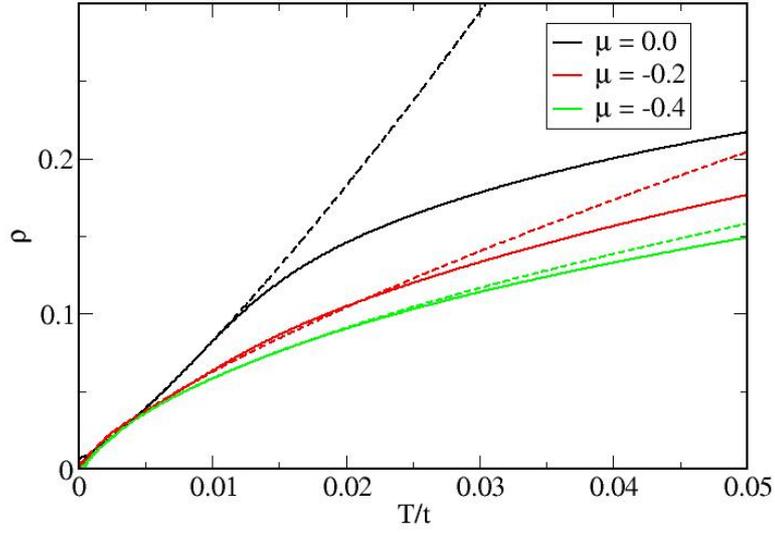

Figure S5. Temperature dependence of the resistivity near a local quantum-critical point (LQCP). The resistivity was estimated in a two-dimensional clean system for three values of the chemical potential $\mu$. The temperature is measured in units of the conduction electron hopping integral $t$ appearing in the dispersion relation $\varepsilon_k = -2t\left(\cos k_x + \cos k_y\right) - \mu$. The numerically raw data (solid lines) are fit at low temperatures by $\rho(T) = a + bT^\gamma$ (dashed lines), with the power exponent $\gamma$ being about 1.23, 0.79, 0.52 corresponding to $\mu$=0, -0.2, -0.4. Tuning the chemical potential down to be more negative adjusts the conduction-electron part of the Fermi surface to be more isotropic, for which the resistivity becomes more sub-linear.

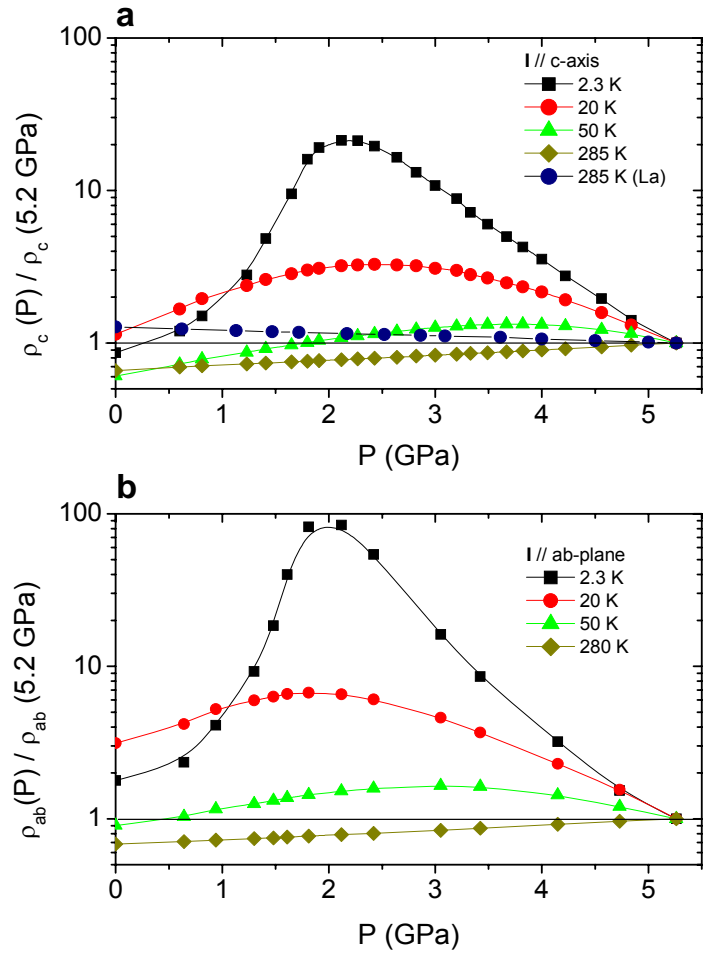

Figure S6. Representative isothermal plots of normalized resistivity of CeRhIn$_5$. **a** Current along the c-axis. At low temperatures, this normalized plot peaks in the pressure range where the superconducting transition temperature is a maximum, as reflected in the colour map given in Fig. 3 of the text. For reference, $\rho(P)/\rho(5.2 GPa)$ at 285 K also is plotted for LaRhIn$_5$, a non-magnetic analog of CeRhIn$_5$. With increasing pressure, the resistivity of LaRhIn$_5$ monotonically decreases because of the increased overlap between adjacent orbitals. The resistivity of CeRhIn$_5$ at 285 K, in contrast, monotonically increases with pressure due to increasing Kondo scattering between localized 4f and conduction electrons. **b** Current perpendicular to the c-axis. The normalized plots peak in the same pressure range as that for current along the c-axis at low temperatures and the relative ratio, $\rho_{ab}(P)/\rho_{ab}(5.2\ GPa)$, is 84 at 2.3 K.